\documentclass[12pt]{article}
\pdfoutput=1
\usepackage{cite}
\usepackage{color}
\usepackage{amsmath}
\usepackage{amsfonts}
\usepackage{amssymb}
\usepackage{caption}
\usepackage{graphicx}
\usepackage{slashed}            % for slashed characters in math mode
\usepackage{xspace}				% For spacing after commands

\usepackage[english]{babel}
\usepackage[applemac]{inputenc}

\usepackage{hyperref}

\def\be{\begin{equation}}
\def\ee{\end{equation}}
\def\ba{\begin{eqnarray}}
\def\ea{\end{eqnarray}}
\newcommand{ \sla }[1]{\setbox0=\hbox{$#1$}         % set a box for #1
   \dimen0=\wd0                                     % and get its size
   \setbox1=\hbox{/} \dimen1=\wd1                   % get size of /
   \ifdim\dimen0>\dimen1                            % #1 is bigger
      \rlap{\hbox to \dimen0{\hfil/\hfil}}          % so center / in box
      #1                                            % and print #1
   \else                                            % / is bigger
      \rlap{\hbox to \dimen1{\hfil$#1$\hfil}}       % so center #1
      /                                             % and print /
   \fi}                                             %

\newcommand{\arXiv}[2]{\href{http://arxiv.org/pdf/#1}{{\tt #2/#1}}}
\newcommand{\arXivold}[1]{\href{http://arxiv.org/pdf/#1}{{\tt #1}}}

\newcommand{\beq}{\begin{eqnarray}}
\newcommand{\eeq}{\end{eqnarray}}

\def\lesssim{\mathrel{\mathpalette\vereq<}}

\makeatletter
\def\vereq#1#2{\lower3pt\vbox{\baselineskip1.5pt \lineskip1.5pt
\ialign{$\m@th#1\hfill##\hfil$\crcr#2\crcr\sim\crcr}}}
\makeatother

\begin{document}

\begin{titlepage}
%%%%%%%%%%%%%%%%%%%%%%%%%%%%%%%%%%%%%%%%%%%%%%%%%%%%%%%%%%%%%%%%%%%%%%%%%%%%
\vskip.5cm

\begin{center} %TITLE HERE
{\huge \bf Planck Data and Ultralight Axions  \\ \vspace*{0.3cm} } 
\end{center}

\vskip1cm

\begin{center}
{\bf Csaba Cs\'aki$^{a}$,
Nemanja Kaloper$^{b}$ and
John Terning$^{b}$}
\end{center}

\begin{center} 
% PLACES HERE
$^{a}$ {\it Department of Physics, LEPP, Cornell University, Ithaca, NY 14853} \\

\vspace*{0.1cm}

$^{b}$ {\it Department of Physics, University of California, Davis, CA 95616} \\

\vspace*{0.1cm}

{\tt  
\href{mailto:csaki@cornell.edu}{csaki@cornell.edu}, 
\href{mailto:kaloper@physics.ucdavis.edu}{kaloper@physics.ucdavis.edu}, 
 \href{mailto:jterning@gmail.com}{jterning@gmail.com}}

\end{center}

\vglue 0.3truecm

\centerline{\large\bf Abstract}
\begin{quote}
We examine the effects of photon-axion mixing on the CMB. We show that if there are very underdense regions between us and the last scattering surface which contain coherent magnetic fields (whose strength can be orders of magnitude weaker than the current bounds), then photon-axion mixing can induce observable deviations in 
the CMB spectrum. Specifically, we show that the mixing can give rise to non-thermal spots on the CMB sky.  As an example we consider 
the well known CMB cold spot, which according to the Planck data has a weak distortion from a black body spectrum, that can be fit by our model.
While this explanation of the non-thermality in the region of the cold spot is quite intriguing, photon-axion oscillation do not explain the temperature of the cold spot itself.  Nevertheless we demonstrate the possible sensitivity of the CMB to ultralight axions which could be exploited by observers.
\end{quote}

\end{titlepage}

\setcounter{equation}{0} \setcounter{footnote}{0}

The QCD axion remains the best solution of the strong CP problem~\cite{pq,weinberg,wilczek}. The most successful UV completions \cite{arvan} which do not suffer from fine tuning \cite{dine,witten} contain many more ultralight axion-like fields. Only one of their linear combinations will couple to the gluons of the Standard Model, becoming the QCD axion. Another linear combination might couple to the photon, and thus remain extremely light. Such an ultralight axion could have cosmological consequences, since it can affect photons coming from distant cosmological sources, including the Cosmic Microwave Background (CMB). 

Previously we considered~\cite{us} the possibility that photon-axion
oscillations \cite{rafsto} in intergalactic magnetic fields 
could contribute to the dimming
of distant supernovae (SNe) \cite{supnov}. This mechanism requires the photon-axion coupling
scale to be $M\sim 4 \cdot 10^{11}$ GeV, which is just above the bounds from SN1987A \cite{raffelt87}, assumes an 
axion mass $m< 10^{-15}$ eV, 
and an intergalactic magnetic field $B\sim 5 \cdot
10^{-9}$ G, with a
domain size of order a Mpc, in agreement with observational bounds
\cite{kronberg,Ade:2013lta}. Further properties of the model were
investigated in~\cite{Csaki:2001jk,Csaki:2003ef,Csaki:2004ha,Csaki:2005vq,jecg,dhuz,grossman,goobar,mrs,husong}.

The current status of this scenario is that while it cannot account for the entire SNe dimming, photon-axion oscillation can still yield a sizeable contribution, contaminating the direct determination of the dark energy equation of state from the supernova Hubble diagram by as much as 
several tens of percents \cite{Csaki:2004ha,mrs}, and making the observed parameter $w = p/\rho$ look more negative than it really is.  
Further, the most precise CMB measurements to date made by the Planck collaboration \cite{Ade:2013lta} have pointed to curious small discrepancies between CMB  and SNe data. Specifically the Planck collaboration determined the fraction of the dark energy to be
$\Omega_\Lambda=0.69 \pm 0.02$ which is smaller than previously obtained by fitting to the dimming \cite{supnov} of 
distant SNe. For example the SNLS combined sample \cite{Ade:2013lta,Conley:2011ku} gives $\Omega_\Lambda=0.77\pm0.05$.
The tension between the different measurement techniques suggests that there may be some as yet unidentified systematic effect involved.
If this survives further (ongoing) scrutiny, it might imply that the dimming of SNe is not only due to the geometry of the universe (via cosmic acceleration), but also some additional non-geometric effect, such as photon-axion mixing. 
Thus it is interesting to look for other signatures of photon-axion mixing to either constrain the couplings or to identify possible glimpses of new axion physics in the sky. One possibility is to search for signatures of photon-axion mixing in the CMB. This has already been considered in \cite{mrs,husong}, where the bounds on photon-axion mixing from non-thermal distortions of the CMB spectrum \cite{smoot} averaged over the whole sky (i.e., the `monopole' in the expansion) have been derived. Nevertheless these bounds still allow for the possibility that small distortions, of the order of observed temperature anisotropies, may be present on some parts of the sky. We will explore this possibility in this note.

Specifically, we will demonstrate that under certain circumstances photon-axion mixing can give rise to non-thermal patches in the CMB sky. This can happen if along the line of sight between us and a section of the surface of last scattering there is a region in space which is very underdense, but also has a nonnegligible magnetic field inside it. Such voids are known to exist in the matter distribution at very low redshifts, arising naturally in the process of structure formation \cite{dubinski}. In fact, they have been invoked \cite{inosilk,Rudnick:2007kw} as a possible partial explanation of the observed CMB cold spot, seen in both the WMAP and Planck data \cite{coldspot}. The idea is that the photons passing through the void between us and the cold spot are cooled down due to the void evolution, affecting the photons through the integrated Sachs-Wolfe effect. In order to explain the cold spot through the integrated Sachs-Wolfe effect the void would have to be extremely large, around 140 Mpc across.
Such large voids would have been seen in large structure surveys, and subsequent scrutiny \cite{nobigvoids} did not reveal them. On the other hand, the current data are not good enough to exclude voids smaller than 
about $50$ MPc, which will be the upper limit on the size of voids we will be considering here.

In this paper we will not try to explain the origin of the CMB cold spot, but will instead explore the origin of the observed weak distortion of the CMB spectrum from a
thermal distribution at the cold spot. We will show that if there is a void along the line of sight smaller than the exclusion bounds mentioned above, and if it is permeated by a coherent magnetic field a couple of orders of magnitude weaker than the current bounds, then photon-axion oscillation could cause the observed distortions of the CMB spectrum. This intriguing agreement  shows the potential sensitivity of the CMB to very light axions. We believe that it should  motivate further and more precise analyses of the 
data to develop better constraints on weakly coupled light particles, or potentially even provide indirect  evidence for their existence.

Let us first review the physics of photon-axion conversions. In a region of constant homogeneous magnetic field $\vec B$, a photon whose electric field is parallel to the background $\vec B$ field can transform into an axion due to the  $a \vec B \cdot \vec E/M$ coupling. 
The conversion probability of photons  with energy ${\cal E}$ into axions over a magnetic domain of size $D$ is given by
\beq
P_{\gamma \rightarrow a} = \frac{4 \mu^2 {\cal E}^2}{(\omega_p^2-m^2)^2+4\mu^2
{\cal E}^2}
\sin^2\left[  \frac{\sqrt{(\omega_p^2-m^2)^2 + 4 \mu^2 {\cal E}^2}}{4{\cal E}}
D\right]
\, .
\label{prob}
\eeq
where  $m$ is the axion mass, $\mu= B/M$, and
$\omega_p$ is the plasma frequency (effective photon mass) \cite{rafsto}
\beq
\omega_p^2=4\pi \frac{\alpha \,n_e}{m_e}~,
\eeq
with $n_e$ the electron density,
$m_e$ the electron mass, and $\alpha$ the fine structure constant. The environmental plasma in the region where mixing occurs is the main 
impediment to conversions, since it makes the photon effectively heavy. 
We will assume that $m \ll  \omega_p$ for realistic values of $n_e$.

One can then easily see that the mixing changes qualitatively as a function of the photon energy, since the behavior of 
$P_{\gamma \rightarrow a}({\cal E}) $ has a crossover (assuming $B D \ll M$), when the 
energy passes through the threshold
\beq
{\cal E}_{\rm th} = \frac {\omega_p^2 \,D}{2} = 2\pi \frac{\alpha \,n_e\,D}{m_e}~.
\label{threshold}
\eeq
When the energy is below ${\cal E}_{\rm th}$ the mixing is very suppressed, while when the energy is greater than 
${\cal E}_{\rm th}$ the mixing is maximal, being essentially independent of the photon energy and controlled only by the coherence length of the 
background magnetic field. For energies around ${\cal E}_{\rm th}$, the mixing is
strongly energy (frequency) dependent. 

In cosmological conditions, this means that the behavior can change as a function 
of distance (or redshift), since the background plasma distribution evolves. At large redshifts, after reionization but before the onset of 
significant clumping, where the matter distribution is still uniform, the plasma frequency is $\sim 10^{-14}$ eV \cite{Csaki:2001jk,dhuz}. At these densities, the critical 
energy ${\cal E}_{\rm th}$ may be large (depending on the domain size of the magnetic field), obstructing photon-axion mixing. 
However, at redshifts $\lesssim 1$, where significant structures are formed and the SNe which are used for the determination of dark energy reside, most of the extragalactic space is underdense by a factor of 10 to 30 \cite{vsilk}. It was estimated in ref.~\cite{Csaki:2001jk} that over most of space
at redshifts $z \lesssim 1$ the electron density is at most
$n_e \leq 6 \cdot 10^{-9} {\rm cm}^{-3}$.
With this value of the electron
density the plasma frequency is $\omega_p \leq 3\cdot 10^{-15}$
eV and for a Mpc sized domain, 
${\cal E}_{\rm th}=  0.7 {\rm  eV}$. 

Thus, fortuitously, the plasma-generated effective photon mass
has the right magnitude to allow for strong mixing of the axion with optical photons, while suppressing photon-axion mixing for sub eV photons so there is only a tiny ($\sim 10^{-8}$) effect on CMB photons. Clearly, for this to work the axion mass should be smaller than the plasma frequency in the relevant regime, which is precisely why we chose the axion mass scale $ \lesssim 10^{-15}$ eV in \cite{us}. In fact, thanks to plasma, we can even relax this, allowing the axion mass to be significantly smaller without affecting the SNe dimming or any of the bounds that come from it, since at larger redshifts the plasma frequency comes in to suppress the mixing. This is a key ingredient for the discussion which follows, where we will consider the effects of localized under densities on the CMB in regions of space where the free electron density is extremely low. If the axion mass is very small, then the mixing effects can reach interesting levels even at larger redshifts and lower photon energies. 

When considering the effect of the photon-axion conversion effect on SNe in \cite{us}, we had to account for the variation of the magnetic fields along the line of sight. That was necessary because typical SNe affected by photon-axion conversion are far away, at distances ranging up to ${\cal O}(1)$ of the Hubble length. On the other hand, the coherence length of nano-Gauss magnetic fields is bounded to be $\lesssim$ MPc. This means that between us and a typical SNe there are many, up to ${\cal O}(1000)$, coherent magnetic domains, with the magnetic field orientation changing randomly from one to another. The precise analysis of what happens for the case with many randomly oriented magnetic domains \cite{us,grossman} shows that the intensity of the photon beam decays exponentially, according to the law $I = (2 + e^{-y/L}) I(0)/3$, where $L \simeq \frac{8M^2}{D B^2}$ and $D \simeq $ Mpc is the magnetic domain size. Carefully accounting for this effect one finds the following bottom line for the effect of photon-axion mixing on SNe: it can significantly influence the luminosity-distance relationship of SNe used to determine the nature of the dark energy dominating the universe. This can affect the effective equation of state parameter $w = p/\rho$ by as much as $10-20\%$ 
\cite{Csaki:2004ha,mrs,husong}, while satisfying other cosmological bounds. The natural question one should then ask is if there are any other signatures of such an effect.

In what follows we will consider the possibility that photon-axion mixing can lead to a localized depletion of CMB photons on isolated sections of the sky, leading to 
frequency-dependent brightness variations.
This would help explain any non-thermal CMB anomalies, such as the 
weak distortions of the Planck distribution observed by the Planck satellite in the CMB cold spot.
%It has been proposed that the cold spot, which has been seen in both the WMAP and Planck data \cite{coldspot} could be explained by the existence of a void \cite{inosilk,Rudnick:2007kw} in the direction of the cold spot. However, in order to explain the cold spot through the integrated Sachs-Wolfe effect the void would have to be extremely large, around 140 Mpc across. Subsequent analysis has constrained the size of any void in that direction to be less that 50 Mpc \cite{nobigvoids}, and so the ISW effect in such realistic voids would not suffice to explain the temperature deviation of the cold spot from the average temperature across the sky. In this paper we will not try to explain the of the origin of the CMB cold spot itself. Rather we will focus on the apparent small deviations from a thermal spectrum  observed near the cold spot. 

\begin{figure}[thb]
\begin{center}
\includegraphics[width=5in]{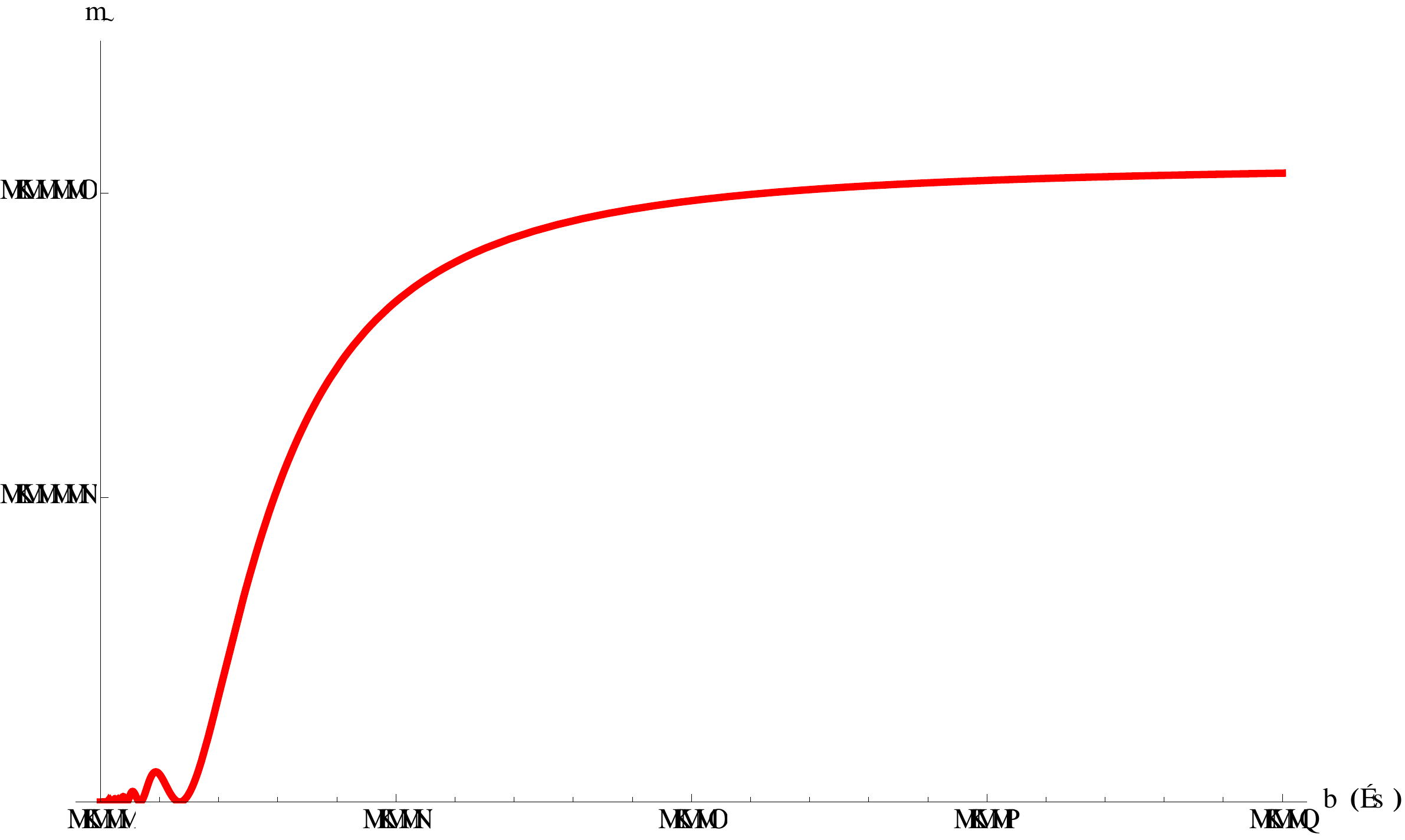}
\parbox{5in}{\caption{The photon oscillation probability for
a void of size 25 Mpc versus photon energy with $\omega_p =  3 \cdot 10^{-17}$ eV and $B = 5 \cdot
10^{-11}$ G. The CMB photons correspond to energies $< 0.0015$ eV.
\label{fig:energy}}}
\end{center}
\end{figure}

The effect of photon depletion due to photon-axion mixing can enhance the temperature variation of a section of the sky due to an inhomogeneity along the line of sight, and this can be frequency dependent. While generically there is almost no effect on CMB photons, as we explained above, localized effects could occur for photons propagating through sufficiently diffuse voids of sizes which obey the bounds found by \cite{nobigvoids}. If the plasma density in the void is much smaller than the average value across the sky, in that region of space the energy threshold (\ref{threshold}) for oscillations decreases. In such a region, the rate of photon depletion due to their transitions into axions will be larger for 
higher frequency photons. The number of photons received at higher frequency would be suppressed relative to the number received at lower frequency, distorting the spectrum. This dependence is shown in Fig.~\ref{fig:energy}. Here, we plot the probability for photons to convert into axions using (\ref{prob}) in a void of the size of 25 MPc, where the plasma frequency is a 100 times smaller that in most of intergalactic space. This means, that the void is underdense by about a factor of $10^4$, and clearly needs to be very long lived. The question of whether such large and strongly underdense voids can exist is still an open problem, attracting ongoing interest \cite{dubinski,sheth,hu}. However, there are simulations which suggest that very underdense voids do form, and once they arise, they are automatically long-lived since most of the matter tends to fall into the surrounding overdensities.  Further we take the magnetic field inside the void to be two orders of magnitude below the cosmological bound of $\sim 5 \cdot 10^{-9}$ Gauss, imagining that it comes from the structures surrounding the void.
For such small magnetic fields the current bounds allow domain sizes  much larger than 1 Mpc. We will assume that the magnetic field is coherent in the entire void of 25 Mpc, while the mixing outside the void is very strongly suppressed due to a larger plasma density. The total oscillation affecting the cold spot will be given by (\ref{prob}) applied to the size of the void, rather than the domain averaged expression used for the SNe dimming.
%
%\vskip.3cm
%

In this case, the threshold energy (\ref{threshold}) is about $0.002$ eV, 
and so the mixing of the CMB photons is still strongly suppressed relative to that of the optical photons, but in a strongly frequency-dependent way.  The mixing depends on the square of the 
frequency in this regime.  Therefore, the CMB photons passing through this void would be depleted at a frequency-dependent rate, the higher frequency ones being depleted more. This is consistent with the bounds on the deviations of the CMB spectrum from a black body, as long as the depletion rate does not exceed $10^{-5}$ or so: the limits on the distortions of the CMB spectrum from that of a blackbody 
only exist for the spectrum averaged over the whole sky, and are 
of the order of the observed thermal anisotropies \cite{smoot,mrs,husong}. 
Clearly, to ensure this, one must have the right conditions in the region where the depletion occurs, correlating the scale of the magnetic field and the size of the domain. 
In fact, with our choice of environmental parameters, the mixing is safely below these bounds.
Concretely, the Planck satellite has observed photons with frequencies of 30 GHz, 44 GHz, 70 GHz, 100 GHz, 143 GHz, 217 GHz, 353  GHz, 545 GHz, and 857 GHz, corresponding to 
energies ranging from $1.2 \times 10^{-4}$ eV
%, $1.8 \times  10^{-4}$ eV, $2.9 \times  10^{-4}$ eV, $4.1 \times  10^{-4}$ eV, $5.9 \times  10^{-4}$ eV, \
%$9.0  \times 10^{-4}$ eV, $1.4 \times  10^{-3}$ eV, $2.3  \times 10^{-3}$ eV, and 
to $3.5 \times  10^{-3}$ eV. Since the galactic dust tends to dominate over the CMB signal at higher frequencies, the three highest frequencies are more useful for removing foregrounds than for actually observing the CMB. So we will focus on the exploration of the six lower frequency channels, while bearing in mind that the 217 GHz bin may be less reliable due to instrument systematics \cite{spergel}.

\begin{figure}[ht]
\begin{center}
\includegraphics[width=1.65in]{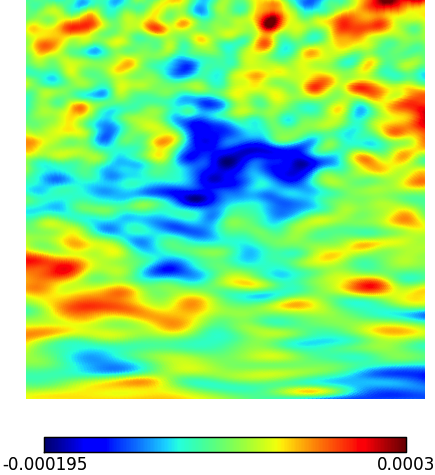}\includegraphics[width=1.65in]{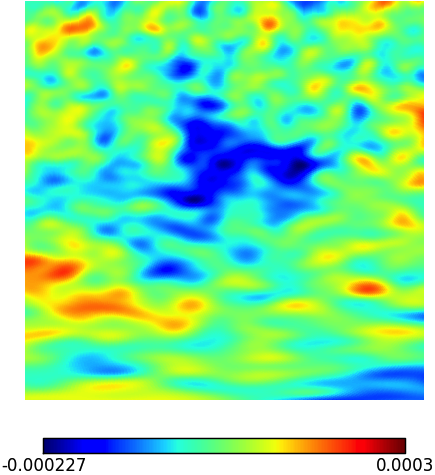}\includegraphics[width=1.65in]{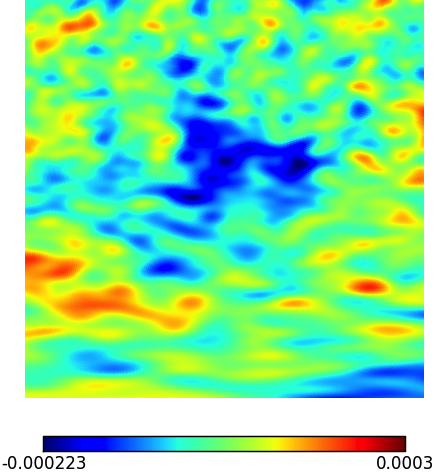}
\end{center}
\begin{center}
\includegraphics[width=1.65in]{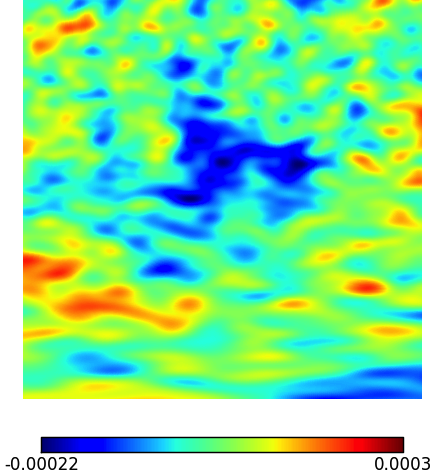}\includegraphics[width=1.65in]{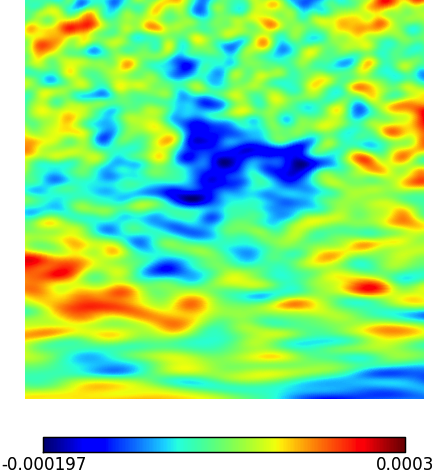}\includegraphics[width=1.65in]{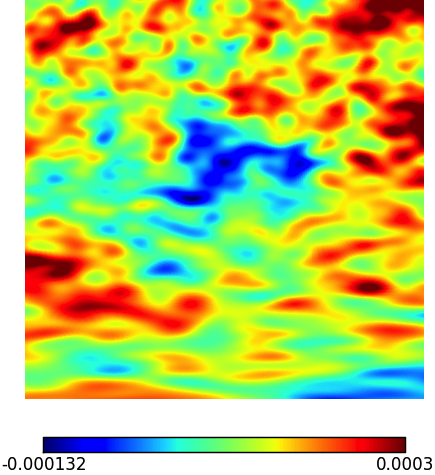}
\end{center}
\begin{center}\parbox{5in}{\caption{Planck data (in Kelvin) from $-170^\circ < \ell < -130^\circ$, and $-80^\circ < b < -40^\circ$  smoothed over $1^\circ$ in the
30 GHz, 44 GHz, 70 GHz, 100 GHz, 143 GHz, and 217 GHz channels. 
\label{fig:planck-cold-spot}}} %%
\end{center}
\end{figure}

A complication in the analysis is that the publicly available raw Planck data is presented as a differential effective black-body temperature for the seven lowest frequencies. So one must properly normalize the temperature, in effect determining the average temperature $T_0$ for the observed region. We have proceeded as follows. For the anisotropy maps corresponding to the lowest 6 frequencies, we smoothed the $\Delta T$ data over $1^\circ$ and then removed the monopole and dipole contributions (excluding low galactic latitudes between $\pm 45^\circ$ from the monopole and dipole fit). Excluding the monopole and dipole yields the local temperature difference, $\Delta T$, from the average CMB temperature of the whole sky, $T_0= 2.7255 \,K$ \cite{Fixsen:2009ug}. We then can extract $\Delta T$ at a particular point in the cold spot ($b= -56.41^\circ$, $\ell = -150.33^\circ$), which we want to explore, 
ignoring the cause of the overall temperature drop. 
We also 
smooth the covariance maps for each frequency over $1^\circ$, extract the covariance at the same spot, take the square root , and divide by the square root of number of pixels in a $1^\circ$ circle. (There are 239 pixels per $1^\circ$ circle in the Low Frequency Instrument data (30 GHz, 44 GHz, 70 GHz), and 958 in the High Frequency Instrument data.) This yields the CMB maps of temperature anisotropies around the cold spot given in 
Fig.~\ref{fig:planck-cold-spot}.
Finally, we compute the intensity of the CMB distribution as a function of frequency, using the local temperature at the cold spot $T_0+\Delta T$ which we obtained by averaging over the six lower frequency channels from Planck, using the Planck distribution, as per the theoretical expectation of having photons in a black body distribution:
\beq
I(f, T) = \frac{2 h}{c^3 }  \frac{f^3}{ (e^{   h f/   k  T} - 1)}~.
\label{bbody}
\eeq
Computing $I(f,T)$ for the six lowest frequency Planck channels yields the theoretical benchmarks to compare to the measured CMB intensities. 

\vskip.3cm

Let us now consider the quantitative aspects of the analysis. 
\begin{figure}[thb]
\begin{center}
\includegraphics[width=5in]{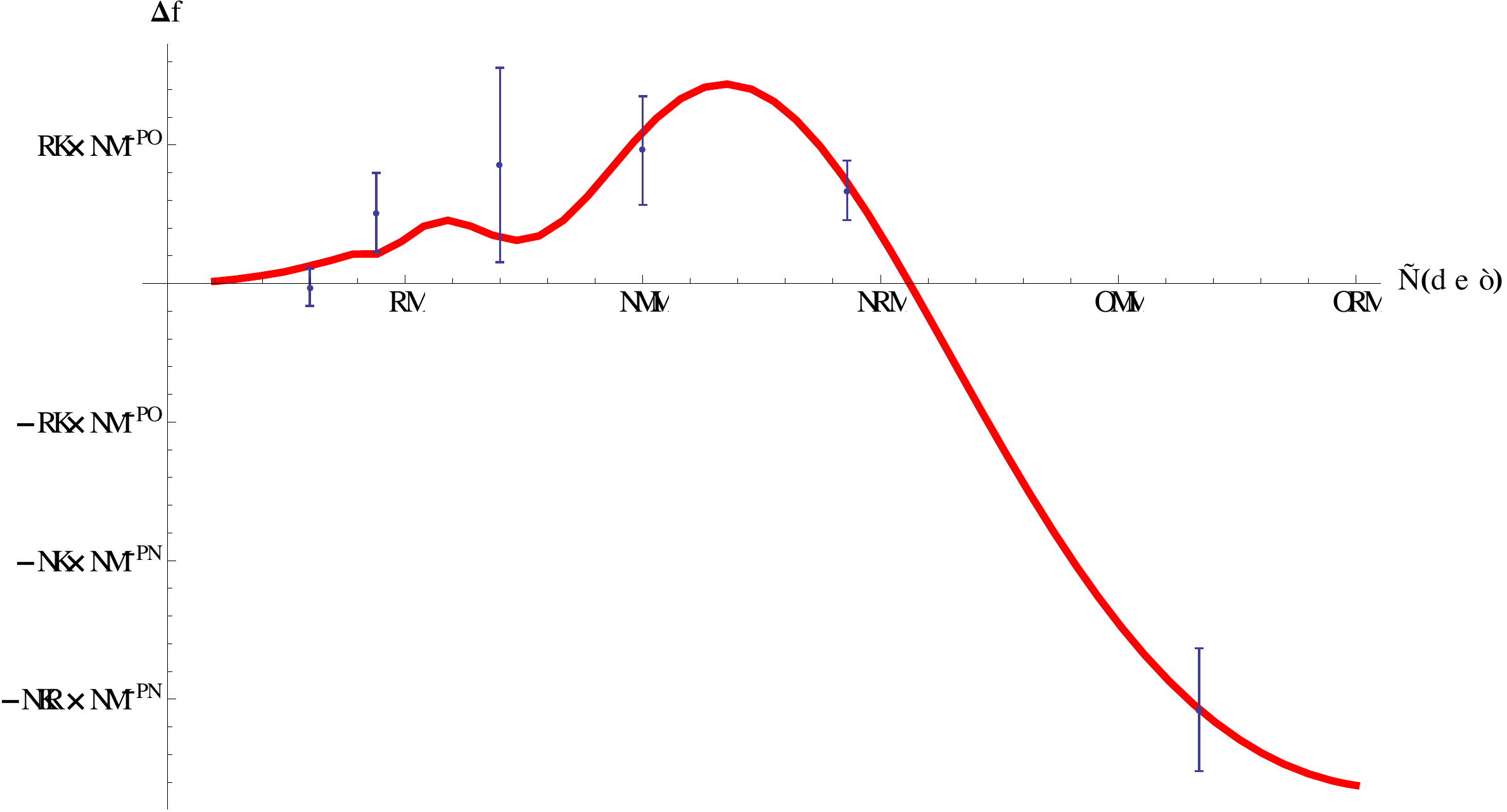}
\parbox{5in}{\caption{Intensity difference $\Delta I$, in units of Js/m$^3$,
versus photon frequency in GHz. The data points correspond to the residual intensity difference $\Delta I$ between the Planck distribution for the local temperature of the cold spot $T_{fit}=T_0-254\, \mu K$. The actual intensities measured by the Planck satellite in the six lower frequency bins. 
The solid line is a fit to the data using a void of size 25 Mpc,  with   plasma frequency of $ 3.9 \cdot 10^{-17}$ eV, a magnetic field $B = 8.6 \cdot
10^{-11}$ G, and with a local temperature 7.5\,$\mu K$ above $T_{fit}$.
\label{fig:fit2}}}
\end{center}
\end{figure}
The six lowest frequency yield the best fit temperature, $T_{fit}$,  that is 254 $\mu K$ below $T_0$ at the cold spot. Using this to normalize the distribution (\ref{bbody}), we plot the intensity residuals with respect to $I(f,T_{fit})$ (the differences between the observed CMB intensities in the six lower frequency channels and $I(f,T_{fit})$) as shown in Fig.~\ref{fig:fit2}.  The plot clearly shows that the black-body hypothesis is a rather poor fit to the Planck data in the cold spot. Note, that there are still uncertainties due to the 217 GHz bin, but even without it, the observed spectrum appears to deviate from thermal at a level consistent with $\Delta T/T \simeq 10^{-6}$. For example, here we have not removed a dust component from the 217 GHz channel, but this would only reduce the intensity, which would yield an even poorer fit; further there may be unknown instrumental systematics. Nevertheless, with the current data the bottom line seems to be that these six bins suggest some distortion of the observed spectrum from a black body.

We {\it can} improve the fit by including photon-axion mixing. A very good fit is provided by inserting a void along the line of sight between us and the cold spot, and slightly altering the void parameters from those mentioned earlier, in the context of  Fig.~\ref{fig:energy}. Taking a slightly higher plasma frequency, $\omega_p = 3.9 \cdot 10^{-17}$ eV (which may be more realistic, given the size of the void and the scale of under density required), and a magnetic field in the void of magnitude $B = 8.6 \cdot
10^{-11}$~G, we plot the results in Fig.~\ref{fig:fit2}. The solid curve shows the residual between $I(f,T_{fit})$ and $I(f,T)P_{\gamma\rightarrow\gamma}$ where $P_{\gamma\rightarrow\gamma}= 1-P_{\gamma\rightarrow a}$, where $T$ is 7.5 $\mu$K warmer than $T_{fit}$, that is, photon-axion mixing does contribute a modest amount to the total cooling. Clearly, the fit looks intriguing. Further checks could be made by comparing the fit including photon-axion mixing with the higher frequency bins, if they could be reliably cleaned from the dust effects and other systematics. 
A future analysis by the Planck collaboration may be warranted to test this prediction.

\begin{figure}
\begin{center}
\includegraphics[width=2.5in]{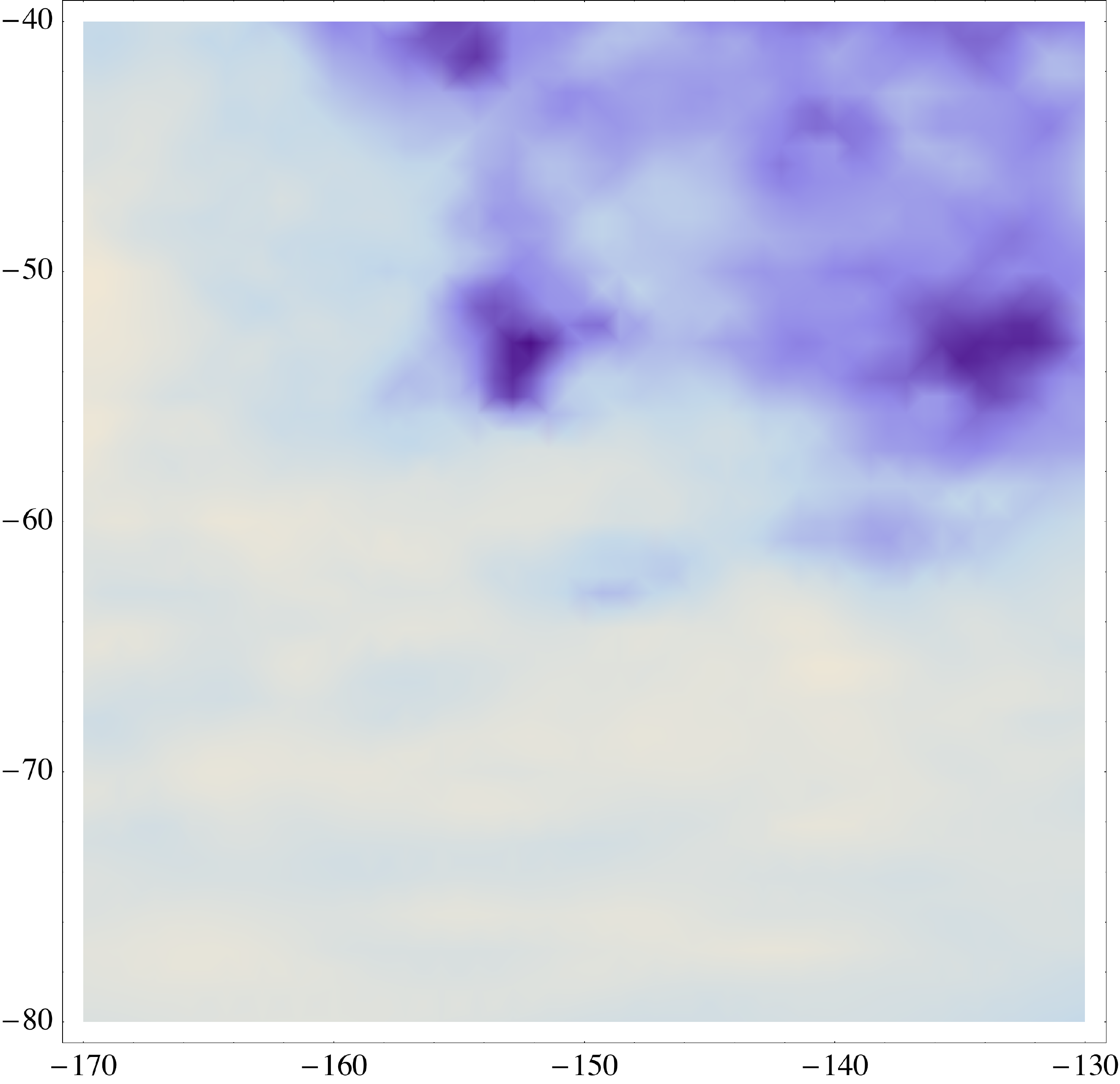}\end{center}
\begin{center}\parbox{5in}{\caption{Temperature differences between the 143 GHz, and 217 GHz channels in the Planck data from $-170^\circ < \ell < -130^\circ$, and $-80^\circ < b < -40^\circ$  smoothed over $1^\circ$. We can see that the cold spot is co-located with a non-thermal spot, and that there are two additional non-thermal spots nearby.
\label{fig:non-thermal}}} %%
\end{center}
\end{figure}
As the last data test, we compare the temperature difference between the 143 GHz and 217 GHz channels around the region of the cold-spot in Fig.~\ref{fig:non-thermal}. If the two channels both fit the black body distribution, the temperatures extracted from the measured intensities would have coincided. In contrast, the spectral distortion generated by photon-axion mixing would produce a non-thermal correction. This seems to be borne out by the data: a non-thermal region shows up on top of the location of the cold-spot as well as some other nearby non-thermal spots. 

While we find the results of this analysis rather intriguing, it is clearly premature to claim that it provides evidence for the existence of an ultralight axion
mixing with the photon. However, the existence of such an axion could help explain some cosmological `blips', such as the non-thermality of the cold spot explored in this paper, and the apparent preference of data for a more negative equation of state $w < -1$ \cite{panstarr,shahut} --- which, while weak, seems more curious in light of the most recent data from cosmic observations. Further, such an ultralight axion could also provide a bit of extra relativistic matter in the universe, 
which might be favored by some recent data analyses \cite{Ade:2013lta,nu}. 
However, the data is still insufficiently accurate for one to reach a definite conclusion. Nevertheless, our analysis clearly demonstrates the possibility of finding imprints of photon-axion mixing with more systematic investigation of the CMB data. A refined cleaning of the data from noise, either instrument-induced or astrophysical, with the photon-axion effect kept in mind to make sure that it is not `cleansed away', seems warranted. Further, if an axion in the relevant window of parameters is really there, and is responsible for the non-thermality of the cold spot in the CMB, then our analysis predicts the existence of a somewhat large and very underdense void. If this is true, one should look for other similar voids in a more precise CMB analysis. At the very least, a better understanding of voids may yield new bounds on physics beyond the Standard Model from future cosmic observations.

%%%%%%%%%%%%%%%%%%%%%%%%%%%%%%%%%%%%%%%%%%%%%%%%%%%%%%%%%%%%%%
%%%%%%%%%%%%%%%%%%%%%%%%%%%%%%%%%%%%%%%%%%%%%%%%%%%%%%

\section*{Acknowledgements}
We thank George Efsthatiou, Brent Folin, Andrew Jaffe, Lloyd Knox, Eichiro Komatsu, Marius Millea and especially Guido D'Amico and Olivier Dore 
for useful discussions and correspondence.
C.C. is supported in part by the NSF grant PHY-1316222.  N.K and J.T. are supported by the
Department of Energy under grant DE-FG02-91ER406746.

\end{document}